\documentclass[twocolumn,twocolappendix]{aastex63}

\received{March 10, 2020}
\revised{June 29, 2020}
\accepted{July 1, 2020}
\submitjournal{AJ}

\shorttitle{HCG LF}
\shortauthors{Yamanoi et al.}

\begin{document}

\title{The $-12$ mag dip in the galaxy luminosity function of Hickson Compact Groups\footnote{Based on data collected at Subaru Telescope, which is operated by the National Astronomical Observatory of Japan.} }


\correspondingauthor{Hitomi Yamanoi}
\email{yamanoi.hitomi@gmail.com}

\author{Hitomi Yamanoi}
\affiliation{Center for Information and Communication Technology, Hitotsubashi University, 2-1 Naka, Kunitachi, Tokyo 186-8601, Japan}
\affiliation{Subaru Telescope, National Astronomical Observatory of Japan, 2-21-1 Osawa, Mitaka, Tokyo 181-8588, Japan}

\author{Masafumi Yagi}
\affiliation{Subaru Telescope, National Astronomical Observatory of Japan, 2-21-1 Osawa, Mitaka, Tokyo 181-8588, Japan}
\affiliation{Department of Advanced Sciences, Hosei University, 3-7-2 Kajinocho, Koganei, Tokyo 184-8584, Japan}

\author{Yutaka Komiyama}
\affiliation{Subaru Telescope, National Astronomical Observatory of Japan, 2-21-1 Osawa, Mitaka, Tokyo 181-8588, Japan}
\affiliation{Department of Astronomical Science, School of Physical Sciences, The Graduate University for Advanced Studies (SOKENDAI), 2-21-1 Osawa, Mitaka, Tokyo 181-8588, Japan}

\author{Jin Koda}
\affiliation{Department of Physics and Astronomy, Stony Brook University, Stony Brook, NY 11794-3800, USA}

\begin{abstract}

We present the galaxy luminosity functions (LFs) of four Hickson Compact Groups using image data from the Subaru Hyper Suprime-Cam. A distinct dip appeared in the faint-ends of all the LFs at $M_g\sim-12$. A similar dip was observed in the LFs of the galaxy clusters Coma and Centaurus. However, LFs in the Virgo, Hydra, and the field had flatter slopes and no dips. 
As the relative velocities among galaxies are lower in compact groups
than in clusters, the effect of galaxy-galaxy interactions
would be more significant in compact groups.
The $M_g\sim-12$ dip of compact groups may imply that
frequent galaxy-galaxy interactions would affect the evolution of galaxies, and
the dip in LF could become a boundary between different galaxy populations. 

\end{abstract}


\section{Introduction} \label{sec:intro}

The galaxy luminosity function (LF) is a powerful tool for describing the properties of galaxy density in various environments. 
A comparison between various LFs provides clues to the environmental dependence
of galaxy formation and evolution.
The LFs of galaxy clusters show a distinct upturn in the $-18<M<-16$ magnitude range
(e.g., \citealt{Binggeli88, Yagi02, Parolin03, Popesso05, Mercurio06, Barkhouse07, R&G08, Lan16}), while the LFs in the field environments have flatter slopes to the faint-end (e.g., \citealt{Blanton05}). The upturn of $M\sim-18$ indicates a division of dominant galaxies between giant and dwarf galaxies (e.g., \citealt{Binggeli88, F&S91, T&H02, deLapparent03}). 

Faint-ends ($M>-12$) of the LFs were investigated in the cluster regions in our local neighborhood. 
\citet{T&T02} represented the faint-end slopes of LFs down to $M_R=-10$ in five different environments. These slopes were not as steep as those of the theoretical mass function obtained from the cold dark matter (CDM) model.
\citet{Yamanoi12} found a significant dip of $M_R\sim-13$ in the LF of the Coma cluster and a steep slope at the faint-end of $M_R>-13$. The Centaurus cluster LF of \citet{C&M06} showed a dip at $-14<M_V<-13$, 
and a sign of a dip was seen in the Fornax LF \citep{Hilker03}.
In contrast, the LFs of the Virgo (\citealt{T&H02, Sabatini03, Lieder12, Ferrarese16}) and Hydra clusters (\citealt{Yamanoi07, Misgeld08}) did not show such a dip.
In a lower-density region such as the loose groups, \citet{Trentham05} provided a composite LF of the neighboring four galaxy groups within the Local Group towards $M_R\sim-10$, and confirmed a weak dip of $M_R\sim-11$. The other result also indicated that some loose groups tend to show a deficiency of galaxies at $-12<M_V<-11$ \citep{Muller18}.

The faint-end dip may reflect the presence of physical processes in galaxy formation and/or evolution that are driven by environmental effects.
A large sample of LFs in various environments is required to understand the origin of the faint-end dip.
In this paper, we focus on compact groups, which have several giant
galaxies concentrated in a small region.
Galaxy interactions would have a more significantly impact in compact groups than in galaxy clusters as the relative velocities of galaxies are smaller, and hence,
the effective interaction timescales are longer,
while interactions with hot gas would be weaker.
Previous studies of the LFs of compact groups are summarized in Appendix \ref{app:lfcg}. There are very few studies of the LFs that cover the magnitude ranges fainter than $M\sim-12$.
To investigate the faint-end dip in compact groups,
we studied the faint-ends of LFs with deep images of
compact groups selected from the catalog of
Hickson Compact Groups (HCGs; \citealt{Hickson82}).

\section{Observations} \label{sec:obs}

The HCG fields were observed with the Hyper Suprime-Cam (HSC: \citealt{Kawanomoto18, Furusawa18, Komiyama18, Miyazaki18}) mounted on the Subaru Telescope. The HSC uses 104 science CCDs, which cover $1.5\degr$ field-of-view in diameter with a pixel scale of $0.17''$. The observation was made in the queue mode on 2016 March 8 UT.
The seeing size (the full width at half maximum, FWHM) was $\sim0.8''$. Although the targets discussed in this paper are only a part of all the targets of our project, and the completion of the project depends on future observations, we analyzed the first four fields that were obtained so far.   
The $g$-band images of the four fields (HCG44, HCG59, HCG68, and HCG79) based on the catalog by \citet{Hickson92} were obtained. Their redshifts were found to be less than 0.015, and no other group or cluster is known around the redshifts in these fields. We summarize our targets in Table \ref{tab:obs}. 
We referred to the NASA/IPAC Extragalactic Database (NED) for the
redshift (z) and the distance modulus ($m-M$) corrected to the cosmic microwave background
reference frame, as shown in Table \ref{tab:obs}, with default cosmological parameters
$H_0=73$ km s$^{-1}$ Mpc$^{-1}$, $\Omega_{\rm{matter}}=0.27$ 
and $\Omega_{\rm{vaccum}}=0.73$. We use the AB-magnitude system throughout this paper.

\section{Data Reduction} \label{sec:reduction}
The $g$-band images were reduced using HSC Pipeline \citep{Bosch18} version 4.0.5. 
We used dome flats for flat fielding. 
The data of the Panoramic Survey Telescope and Rapid Response System 1 (Pan-STARRS1) were referred to as astrometric catalogs. 

\begin{figure}
\plotone{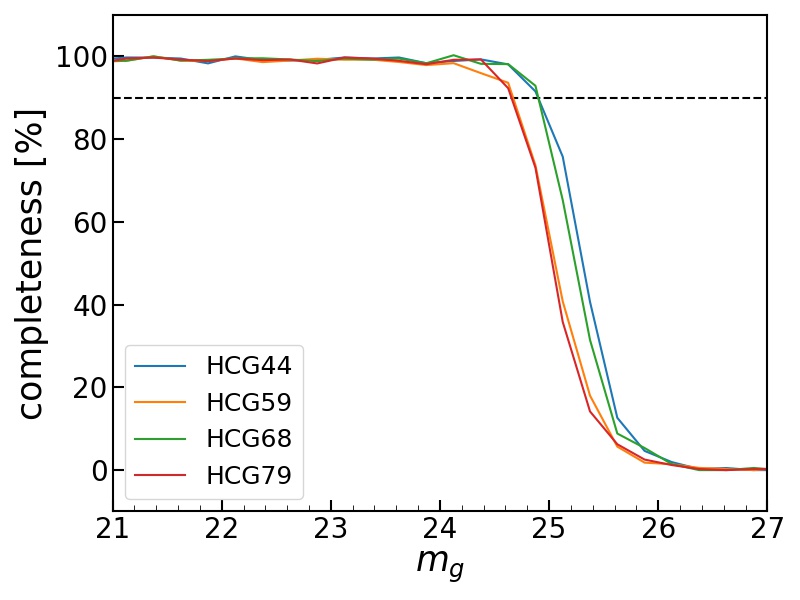}
\caption{
Completeness of extended objects as a function of $g$-band magnitude for each HCG image.
The 90\% completeness is plotted as dashed line.
HCG79 and HCG59 show more than 90\% completeness at $m_g<24.7$, while the 90\% completeness of HCG44 and HCG68 are down to $m_g\sim25$. 
\label{fig:completeness}}
\end{figure}

We detected objects using SExtractor \citep{B&A96} and adopted {\tt MAG\_AUTO} for measuring the magnitude, while excluding the saturated objects.   
The limiting magnitude of point sources was estimated by measuring sky counts with the $2''$-diameter aperture in the reduced image. We applied the Gaussian fitting of the histogram of sky counts and calculated the $5\sigma$ limiting magnitude, which resulted in 
$\sim26$ mag in each image (see Table \ref{tab:obs}). We also estimated the detection completeness of extended sources, creating mock galaxies in the magnitude range of 20.0 -- 28.0 with the Image Reduction and Analysis Facility (IRAF: \citealt{Tody86, Tody93}) task {\tt gallist}. We embedded these galaxies into the object subtracted images, which were obtained as the SExtractor check images of {\tt-OBJECTS}, using {\tt mkobjects}. The galaxies were measured using SExtractor, and the corresponding parameters were adjusted to be the same as the detecting threshold for real objects. We evaluated the detection rates for every 0.25 magnitude bin for the mock galaxies on the entire field of view of the HSC image.
Figure \ref{fig:completeness} indicates that a 90\% completeness fraction was attained  for the extended objects at $m_g=24.7$, which corresponds to $M_g=-9.4$ for the most distant galaxy group (HCG79).
No completeness correction was performed when we evaluated the LFs in the $>90\%$ completeness range. 

The galactic extinctions in $g$-band ($A_g$) in Table \ref{tab:obs} obtained from NED were adopted when we converted magnitudes from the apparent to the absolute values.

\begin{deluxetable*}{cccccccc}
\tablecaption{Observed HCGs\label{tab:obs}}
\tablewidth{0pt}
\tablehead{
\colhead{Field} & \colhead{RA} & \colhead{Dec} & \colhead{z} & \colhead{Exp. Times} & \colhead{$m-M$} & \colhead{$A_g$} & \colhead{$m_{\rm lim}$ ($5\sigma$) } \\
\nocolhead{Field} & \colhead{(J2000.0)} & \colhead{(J2000.0)} & \nocolhead{z} & \colhead{(min)} & \colhead{(mag)} & \colhead{(mag)} & \colhead{(mag)}
}
\decimalcolnumbers
\startdata
HCG44 & 10:20:45.5 & +21:33:37 & 0.0057 & 78 & 31.8 & 0.09 & 26.0 \\
HCG59 & 11:51:01.0 & +12:25:59 & 0.0147 & 60 & 33.9 & 0.11 & 26.3 \\
HCG68 & 13:55:42.4 & +40:05:17 & 0.0086 & 60 & 32.8 & 0.04 & 26.2 \\
HCG79 & 16:01:23.8 & +20:37:11 & 0.0148 & 30 & 33.9 & 0.21 & 26.1 \\
\enddata
\end{deluxetable*}  

\section{Luminosity Function} \label{sec:lf}

\begin{figure}
\plotone{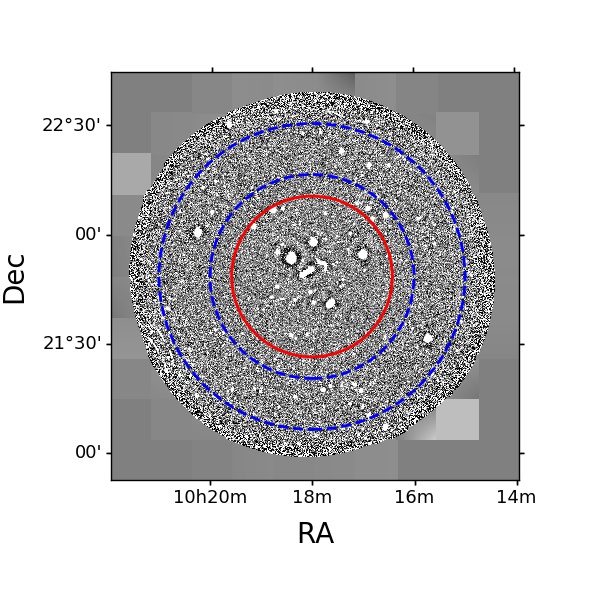}
\caption{
The $g$-band image of HCG44.
The region of HCG is within the red circle with a radius of 150 kpc. The ring region enclosed by the broken blue circles is the control field for the statistical background subtraction.  
\label{fig:image_hcg44}}
\end{figure}

In order to estimate the number of member galaxies, we used the statistical background subtraction method \citep{Bernstein95}. Our HSC image has a field of view of $\sim1.5\degr$ in diameter and is wide enough to cover both the HCG and its outer regions.  
Additionally, we endured that 90\% detection completeness was achieved 
(see Section \ref{sec:reduction}).
We adopted an outer ring area within the range of 10000 -- 15000 pix
($\sim 0.47\degr$ -- $0.71\degr$) from the center of the galaxy group
as a control field (Figure \ref{fig:image_hcg44}), and
counted the number of foreground and background galaxies $N_{\rm cf}$ in the field.
For our fields, we found that the background of each group is represented better with the ring around it (our control field) than with the combined field of control fields of all the groups. The field-to-field variations are relatively large among the HCGs.
More details of the analysis of the background are given in Appendix \ref{app:bg}.  
We adopted the radius of the group region as 150 kpc,
which corresponds to the virial radius of a typical mass of HCG (several $10^{12}M_\sun$).
We then derived the number of galaxies in the group field $N_{\rm gf}$ within a radius of 150 kpc. The number of member galaxies $N_{\rm mb}$ is given by
\begin{equation} \label{eq:lf}
N_{\rm mb}=N_{\rm gf}-kN_{\rm cf},
\end{equation}
where $k$ is the coefficient of the field size correction to the group region. 
Its uncertainty $\sigma_{\rm mb}$ was subjected to the Poisson statistics:  
\begin{equation} \label{eq:lf_error}
\sigma_{\rm mb}=\pm\sqrt{N_{\rm gf}+k{^2}N_{\rm cf}}.
\end{equation}
The subtraction was performed at 1.0 mag intervals.

\begin{figure*}
\plotone{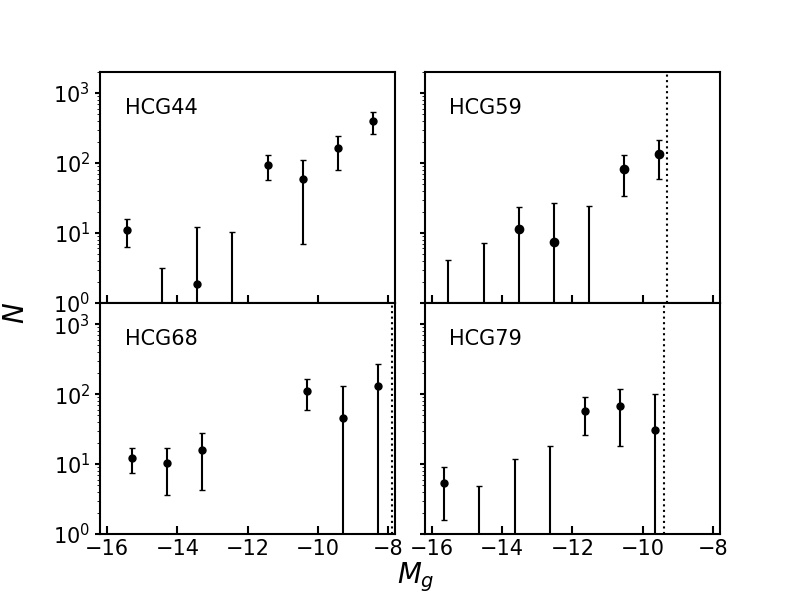}
\caption{
The $g$-band LFs for individual groups of HCG44, HCG59, HCG68 and HCG79.
The error in each magnitude bin is given by Poisson statistics.
The galaxy counts are estimated by the projected number counts within a radius of 150 kpc in the region of the HCGs.
The vertical dotted line indicates the 90\% completeness limit for each HCG distance.
\label{fig:lf_indiv}}
\end{figure*}

\begin{figure}
\plotone{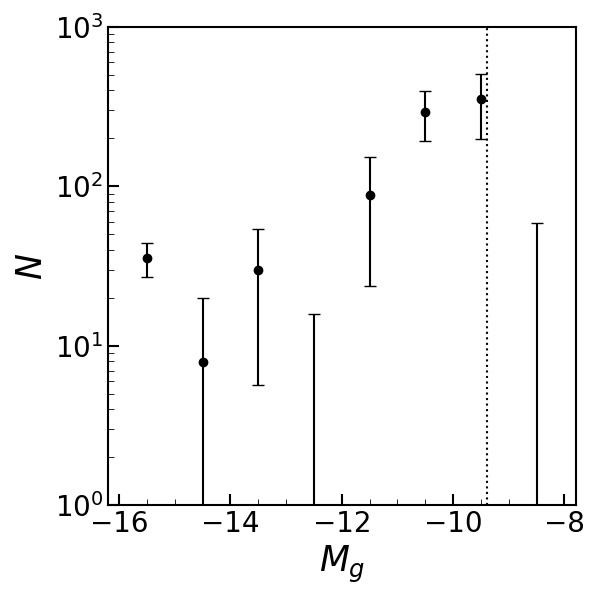}
\caption{
The total LFs of the four HCGs per 1.0 mag interval.
The estimations of the errors and galaxy counts are the same as those in Figure \ref{fig:lf_indiv}. 
The 90\% completeness limit of the most distant compact group in our sample HCG79 is shown by the vertical dotted line.
\label{fig:lf_total}}
\end{figure}

We show the LFs of the HCGs down to $M_g\sim-9$ in Figure \ref{fig:lf_indiv}. 
All groups showed a dip at around $-13<M_g<-12$.
We confirmed that this dip did not come from the trend of the galaxy number counts of the control field in each HCG (Figure \ref{fig:bg_ncount}). No feature which makes the dip of the LF was found. 
The numbers of member galaxies were estimated statistically using the Equation (\ref{eq:lf}).
In this magnitude range, the numbers were consistently low
among all the HCGs and were often negative after background subtraction.
This suggests a statistically significant deficit in group members around this magnitude. 
In order to construct the total of the LFs of all the HCGs, we first converted the apparent magnitude into the absolute magnitude, and computed the number of member galaxies in bins of 1.0 mag for each HCG field. The four LFs were summed up in the same magnitude interval.  
Figure \ref{fig:lf_total} shows the total LF, which is the sum of the four HCGs.

\citet{Krusch06} measured the $B$-band LF of five HCGs (HCG16, 19, 30, 31 and 42
) and found that the LF increased rapidly at $-15<M_B<-13$ and decreased at $M_B>-12$. 
They discussed that this downturn was the cause of the incompleteness according to the detection limit. Our LFs complement the previous result as these attained fainter magnitudes.
The sign of a dip was found at $M_g\sim-15$, and the clear upturn at $M_g\sim-12$ appeared toward the fainter-end, as shown in Figure \ref{fig:lf_total}.
The $-15$ mag dip is similar to that of \citet{Krusch06}, but in our analyses the numbers of galaxies around $-15$ mag are relatively small, and it is difficult to confirm the $-15$ mag dip due to the poor statistics.
In contrast, the intrinsic galaxy population increases toward the fainter magnitude; combined with our fainter limiting magnitude, our analysis includes a sufficient number of galaxies around the $-12$ mag to confirm the dip.

\section{Discussion and Summary} \label{sec:discuss}

\begin{figure}
\plotone{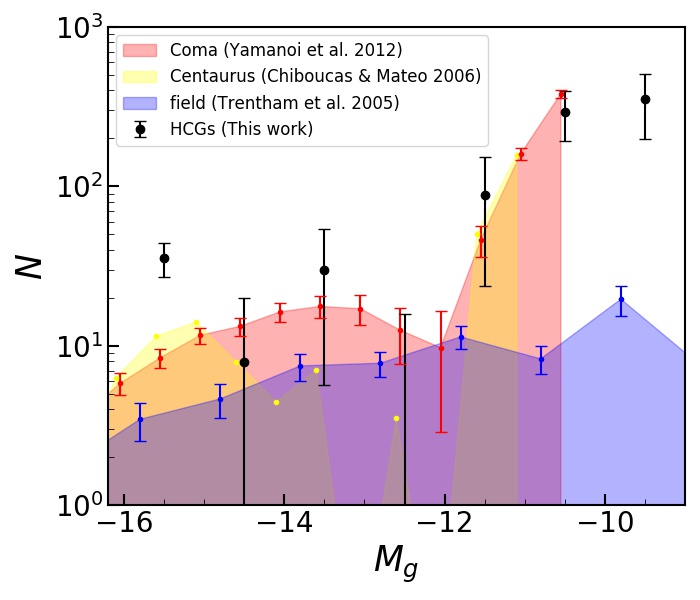}
\caption{
The total LFs of four HCGs and LFs from previous studies.
The vertical scale is arbitrary for LFs of Coma, Centaurus, and field.
These magnitudes are shifted horizontally to adjust to the $g$-band. 
\label{fig:lf_compare}}
\end{figure}

The presence of the $-12$ mag dip in LF may have important implications on the process of galaxy formation.
Environmental dependence, if it exists, would provide an additional clue to understanding the key physical process.
We compared the HCG LF with LFs of other environments.
In Figure \ref{fig:lf_compare}, the LFs of the HCGs, Coma ($B$-band; \citealt{Yamanoi12}), Centaurus ($V$-band; \citealt{C&M06}), and the field ($R$-band; \citealt{Trentham05}) are shown. These LFs from previous literature used various bands. 
We converted these to the $g$-band magnitude by assuming the typical color of galaxies 
to be $(g-r)\sim0.5$.
The number of galaxies $N$ was arbitrarily scaled for comparison.  
The LFs of the Coma and Centaurus showed a significant faint-end dip at $M_g\sim-12$, while the LF of the field did not show a significant faint-end dip. 
The faint-end dip was clearly seen in all four HCGs (see Figure \ref{fig:lf_indiv}).
No dip appeared in the Virgo and Hydra clusters (e.g., \citealt{T&H02, Sabatini03, Lieder12, Ferrarese16, Yamanoi07, Misgeld08}).
The LF of some loose groups of galaxies were found to have a small dip \citep{Trentham05}. 
The loose groups have velocity dispersions similar to those of the compact groups, 
while their densities are much lower. 
The loose groups and HCGs are generally less massive ($\sim10^{12-13}M_\sun$; \citealt{Hickson92}) than the clusters.
Nevertheless, the HCGs and Coma, with a massive cluster of $\sim10^{15}M_\sun$ \citep{Kubo07}, showed a similar, distinct dip.
Hence, we suggest that this faint-end dip is independent of the cluster mass.

Ram-pressure stripping is often discussed as an environmental effect on galaxies.
It removes the gas from dark matter halos and truncates star formation, 
consequently making the galaxies red (e.g., \citealt{F&S80, White91, S&S92}).
However, the ram pressure is effective only in cluster halos with sufficient quantities of hot gas.
Most compact groups have lower characteristic temperatures in X-ray emissions than that of the galaxy clusters \citep{Ponman96}. Hence, the ram-pressure is not the most plausible cause 
of the $-12$ mag dip.
The gravitational potential of the compact groups is not as deep as that of clusters,
and hence the tidal effect due to their potential is unlikely to be efficient.
Thus, the ram-pressure and the tidal effect due to the gravitational potential 
of groups/clusters are unlikely to be the cause of the faint-end dip.

Interactions between multiple galaxies often occur in compact groups
because of their high density. The effect of galaxy-galaxy interactions
is likely to be larger in compact groups because of the lower velocity dispersion of $<230$ km s$^{-1}$ \citep{Hickson92} which causes the typical timescale for interactions to be longer.
On the other hand, rich clusters, such as Coma, have a high dispersion of $\sim1000$ km s$^{-1}$, and the effect of the interactions is relatively weaker.
\citet{Miles04} discussed, with a toy simulation, that the dynamical friction is more efficient in a lower velocity dispersion system, triggering mergers of intermediate-luminosity galaxies into giants around the center, and thus, leading to a dip in a LF.
The galaxy-galaxy interactions are a possible, and more likely cause of the $-12$ mag dip. 

In addition to the observationally motivated discussions given above, we now discuss the dip from the point of view of the hierarchical clustering models of structure formation (i.e., the CDM model).
The theoretical model predicted that the mass function of the dark
matter halo has a steep slope and is monotonic \citep{W&F91, Cole91, Kauffmann93}.
Therefore, the presence of a dip in LF indicates that a certain population of galaxies suffers from some physical changes,
such as disruption, enrichment, and brightening. The dip offers a clue to understanding the physical process that deprives a galaxy population of a specific luminosity.
For example, according to a recent numerical study of dark matter halos,
baryonic feedback is very efficient in dwarf galaxies of
$\sim10^{7-9}M_\sun$ (corresponding to $-14<M<-12$), which can even
change halo mass profiles within these galaxies \citep{B&BK17}.
The more diffuse halos of this galaxy population could be more susceptible to tidal disruption than those of fainter dwarfs.
If this is truly the case, this mechanism could suppress the LF in this luminosity range and may result in a dip in LFs around $M\sim-12$.

Alternatively, the dip of $M_g\sim-12$ may indicate a change in the dominant galaxy population.
Previous studies suggest that the dips of LF are boundaries between two different populations. The clear dip at a brighter magnitude of $M\sim-18$ is widely known, where the dominant population changes from giant galaxies to dwarfs (e.g., \citealt{Binggeli88, F&S91, T&H02, deLapparent03, Miles04}).
Another dip appears at $M\sim-15$ in rich clusters, which is suggested to be a partition between dwarf elliptical and dwarf irregular populations \citep{Wilson97}.  
The LF of \citet{Krusch06} in the HCGs have a similar dip at $M_g\sim-15$.
Several studies on the infrared properties of galaxies in compact groups
have shown a gap between gas-rich and gas-poor populations in an infrared color space, while such a separation is not seen in field galaxies \citep{Johnson07, Walker10, Zucker16}.
These studies suggest a rapid evolution from star-forming to quiescent galaxies in compact groups. 
\citet{Coenda15} compared galaxies in compact groups and loose groups and
found that there were two populations of late-type galaxies in compact groups, one with normal star formation rates and another with low star formation rates. 
If such a partition occurs preferentially along the galaxy mass/luminosity, 
the significant dip of HCGs could be related to the populations of late-type galaxies.

In addition to the physical processes inside the groups and clusters, there is also a possibility of external influence. The dip of a cluster LF may be explained partly by an infalling compact group.
\citet{Cortese06} found a compact group infalling into a dynamically young cluster.
If galaxy clusters were enriched by accreting compact groups,
the LF of cluster galaxies might inherit the dip imprinted in the LF of the original group galaxies. In a previous study of the Coma cluster, \citet{Yamanoi12} compared the LFs of the cluster center and outer regions within the cluster and found a dip of $M_R=-13$ in both regions, similar to those of the HCGs. 
Hence, the dip seems to be independent of the locations within the cluster.
If the dip of a cluster is an inheritance from compact groups, a physical process that is relevant in the groups could also explain the dip in the Coma cluster.
In this case, the compact groups must be the major building blocks of the cluster.
We should, however, note that this idea cannot be applied to all clusters. 
A larger sample is needed to identify the key parameter to the dip.

In summary, we constructed LFs of four neighboring compact groups to $M_g\sim-10$.
All the groups showed a dip at $\sim-12$ mag, which resembles the LFs of several rich clusters. 

\acknowledgments
We thank the anonymous referee for helpful comments.
We acknowledge the support astronomers, the queue team and the archive team of the Subaru Telescope for the observation and the data distribution.
We are grateful for discussion with Kohei Hayashi.
This research has made use of the NASA/IPAC Extragalactic Database (NED),
which is operated by the Jet Propulsion Laboratory, California Institute of Technology,
under contract with the National Aeronautics and Space Administration.
Data analysis was in part carried out on the Multi-wavelength Data Analysis System operated by the Astronomy Data Center (ADC), National Astronomical Observatory of Japan.
YK acknowledges support from JSPS Grant-in-Aid for Scientific Research (JP15K05037) and
MEXT Grant-in-Aid for Scientific Research on Innovative Areas (JP15H05892).
JK acknowledges support from NSF through grant AST-1812847. 
The Pan-STARRS1 Surveys (PS1) and the PS1 public science archive have been made possible through contributions by the Institute for Astronomy, the University of Hawaii, the Pan-STARRS Project Office, the Max-Planck Society and its participating institutes, the Max Planck Institute for Astronomy, Heidelberg and the Max Planck Institute for Extraterrestrial Physics, Garching, The Johns Hopkins University, Durham University, the University of Edinburgh, the Queen's University Belfast, the Harvard-Smithsonian Center for Astrophysics, the Las Cumbres Observatory Global Telescope Network Incorporated, the National Central University of Taiwan, the Space Telescope Science Institute, the National Aeronautics and Space Administration under Grant No. NNX08AR22G issued through the Planetary Science Division of the NASA Science Mission Directorate, the National Science Foundation Grant No. AST-1238877, the University of Maryland, Eotvos Lorand University (ELTE), the Los Alamos National Laboratory, and the Gordon and Betty Moore Foundation.

\vspace{5mm}
\facilities{Subaru (HSC)}

\software{SExtractor \citep{B&A96}, HSC Pipeline (v4.0.5; \citealt{Bosch18}), IRAF \citep{Tody86, Tody93}}

\clearpage 

\appendix
\restartappendixnumbering

\section{Previous Studies of Compact Group LFs} \label{app:lfcg}

We summarized the LFs of compact groups in the previous studies in Table \ref{tab:lfcg}.
We re-calculated the limiting magnitude of each LF study using $H_0=73$ km s$^{-1}$ Mpc$^{-1}$.
All the studies in Table \ref{tab:lfcg}, except \citet{Krusch06}, provided explicit $H_0$ ($=100h$ km s$^{-1}$ Mpc$^{-1}$).
We assumed that \citet{Krusch06} adopted $h=1$ as it consistently explains the apparent magnitude of bright members in their sample and the overall shape of their LF.

\begin{deluxetable*}{lccl}
\tablecaption{LFs of Compact Groups\label{tab:lfcg}}
\tablewidth{0pt}
\tablehead{
\colhead{Reference} & \colhead{\# of groups} & \colhead{z} & \colhead{LF limit} 
}
\decimalcolnumbers
\startdata
\citet{HT80} & 10 & $z<0.017$ & $M_B=-17.2$  \\
\citet{MH91} & 68 & $0.005<z<0.139$ & $M_B=-16.7$  \\
\citet{Ribeiro94} & 22 & $z<0.03$ & $M_B=-15.2$  \\
\citet{SR94} & 68 & $z\leq0.03$ & $M_{B_T}=-15.6$  \\
\citet{Zepf97} & 17 & $z\leq0.03$ & $M_B=-15.1$  \\
\citet{Hunsberger98} & 39 & $z\leq0.05$ & $M_R=-14.2$  \\
\citet{Muriel98} & 66 & $0.009\leq z\leq0.032$ & $M_{b_J}=-16.7$  \\
\citet{ZM00} & 6 & $0.009<z<0.026$ & $M_R=-16.7$  \\
\citet{Zandivarez03} & 451 & $0.003\leq z\leq0.25$ & $M_{b_J}=-15.4$  \\
\citet{KF04} & 69 & $0.008\leq z\leq0.018$ & $M_B=-17.2$  \\
\citet{Miles04} & 25 & $z\leq0.016$ & $M_B=-12.9$  \\
\citet{Krusch06} & 5 & $z\leq0.015$ & $M_B=-10.9$  \\
\citet{Coenda12} & 846 & $0.06\leq z\leq0.18$ & $M_{0.1_r}=-18.9$  \\
This work & 4 & $0.005<z<0.015$ & $M_g=-9.4$  \\
\enddata
\tablecomments{Each magnitude limit is converted with $H_0=73$ km s$^{-1}$ Mpc$^{-1}$.}
\end{deluxetable*}  

\section{Variations of Control Field} \label{app:bg}

\begin{figure}
\plotone{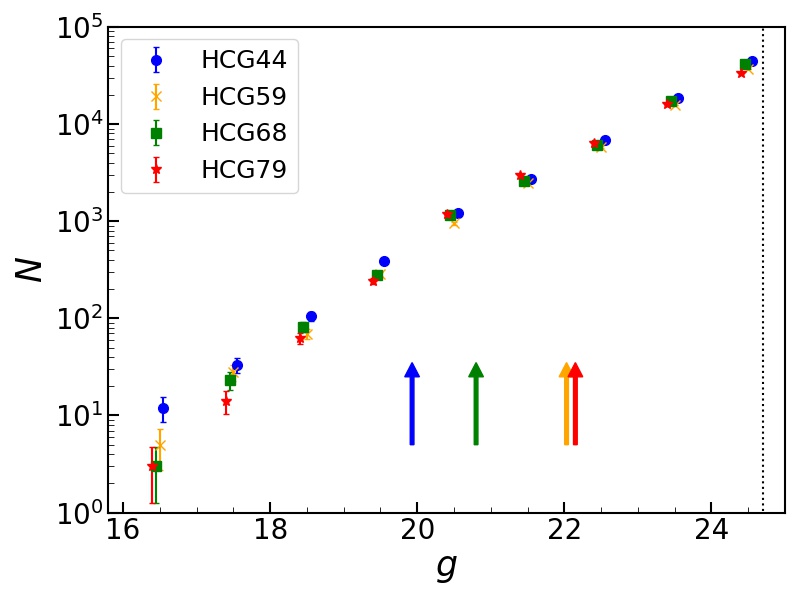}
\caption{
The galaxy number counts of the control field of each HCG.
All the points except HCG59 are shifted slightly in the horizontal direction to avoid an overlap of the points.
The vertical dotted lines indicate the 90\% completeness limit for the
most distant galaxy group (HCG79).
The arrows show the locations of the absolute magnitude of $-12$ at the distance of each HCG.
\label{fig:bg_ncount}}
\end{figure}

For the statistical background subtraction in Section \ref{sec:lf}, we defined a control field as an area within an annulus of 10000 -- 15000 pix from the center of each galaxy group  (see Figure \ref{fig:image_hcg44}). 
In Figure \ref{fig:bg_ncount}, we showed the galaxy number counts of the control field of each groups. We confirmed that no bump nor notable feature was seen in the galaxy number counts of the control field of each HCG. This indicates that 
the dip of $M_g\sim-12$ in the LFs does not come from their control fields. For further inspections of the field-to-field variations around HCGs,
we performed the following tests to confirm that this locally-defined control field represents the background population well around the $-12$ mag dip of LF.

\begin{figure*}
\plotone{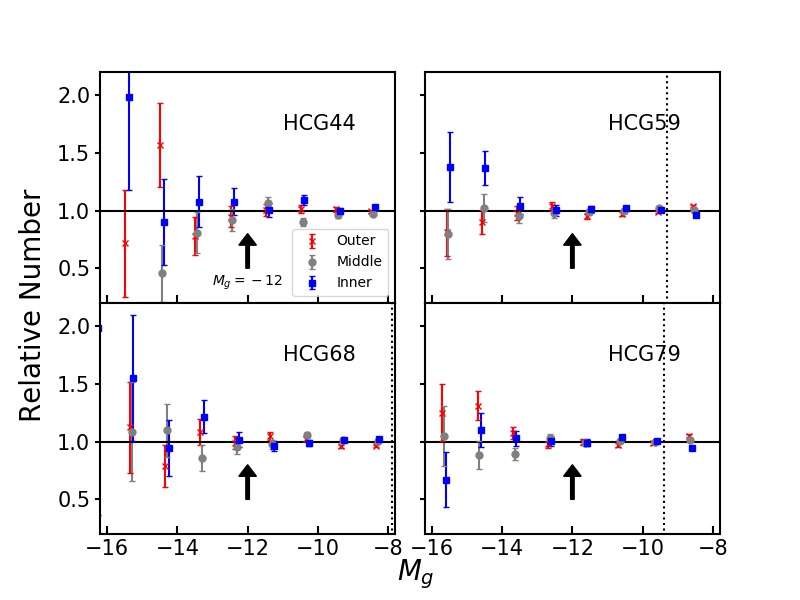}
\caption{
Comparisons of galaxy number counts between the entire control field and three narrower annuli within the control field. The entire control field spans the radial range of 10000 -- 15000 pix, and the inner, middle, and outer annuli represents the ranges of 10000 -- 11000, 12000 -- 13000, and 14000 -- 15000 pix.
The vertical dotted lines indicate the 90\% completeness limit.
The points of inner and outer rings are shifted slightly in the horizontal direction to avoid an overlap of the points.
The arrows show the locations of the absolute magnitude of $-12$ at the distances of the HCGs.
We find no significant radial trend around $M_g\sim-12$. 
\label{fig:bg_org}}
\end{figure*}

First, we investigated potential radial variations within the control field of each HCG.
We divided the control field into three narrower annuli, i.e., inner (10000 -- 11000 pix), middle (12000 -- 13000 pix), and outer (14000 -- 15000 pix) rings.
Figure \ref{fig:bg_org} shows the relative galaxy counts in each ring with respect to those of the control field.
Although there are scatters in the bright magnitude range due to poor statistics, they converge to 1 in the fainter ranges of $M_g>-14$.
Thus, the radial dependence within the control field is not significant, and we can assume that the variation is small between each HCG region and the corresponding control field.

Second, we compared the control field of each HCG (local control field) with the combined field of control fields of the 4 HCGs (global control field).
The four panels show the deviations of the local fields from the global one, and the scatters are large.
We therefore concluded that the local control fields, the areas immediately next to the HCG fields, better represent the local background for the statistical correction.

\begin{figure*}
\plotone{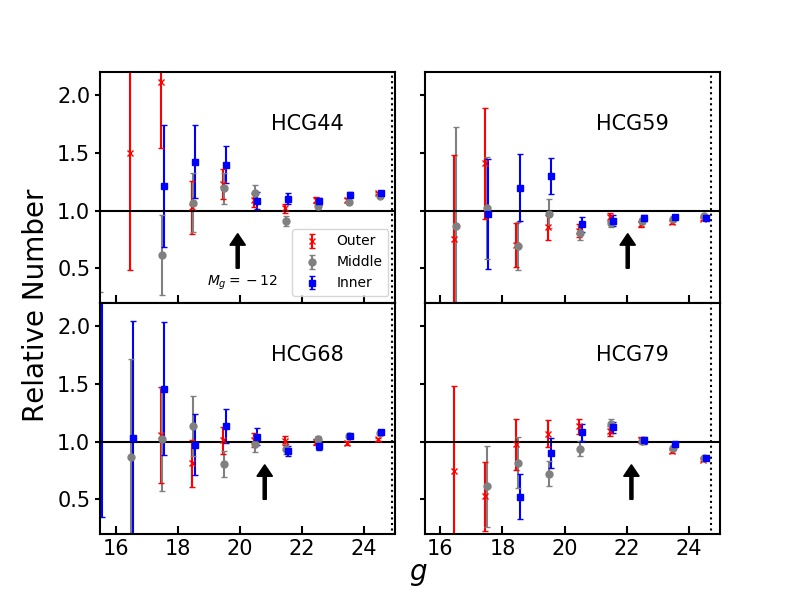}
\caption{
The same as Figure \ref{fig:bg_org}, but between the combined field of the 4 control fields and each control field.
The field-to-field variations are large.
\label{fig:bg_comb}}
\end{figure*}


\bibliographystyle{aasjournal}

\end{document}